\documentclass[sts]{imsart}
\usepackage{graphicx}
\usepackage{geometry}
\usepackage{amsmath,amsfonts,bezier,amstext,amsthm,amssymb}
\usepackage{amsmath,amsfonts,amssymb,amscd,bezier,amstext,makeidx}
\usepackage{natbib}
\usepackage[latin1]{inputenc}
\usepackage{makeidx}
\usepackage{hyperref}
\usepackage{longtable}

\arxiv{math.PR/0000000}

\begin{document}

\begin{frontmatter}

\title{Bayesian model averaging: A systematic review and conceptual classification \protect\thanksref{T1}}
\runtitle{BMA: Review and classification}
\thankstext{T1}{AMS 1991 subject classifications: 62F15; 62C10; key words and phrases:  Bayesian Model Averaging; Systematic review; Conceptual classification scheme; Qualitative content analysis}

\begin{aug}
  \author{Tiago M. Fragoso \thanksref{t2}\ead[label=e1]{fragoso@ime.usp.br}}
  \and
  \author{Francisco Louzada Neto\ead[label=e2]{louzada@icmc.usp.br}}

  \thankstext{t2}{To whom correspondence should be addressed. This work was partially funded by the Brazilian's Ministry of Education Coordena\c{c}\~{a}o de Aperfei\c{c}oamento de Pessoal de N\'{i}vel Superior (CAPES) through a postdoctoral PNPD research grant}

  \runauthor{Fragoso and Louzada}

  \affiliation{Departamento de Matem\'{a}tica Aplicada e Estat\'{i}stica -  Instituto de Ci\^{e}ncias Matem\'{a}ticas e de Computa\c{c}\~{a}o, Universidade de S\~{a}o Paulo}

  \address{Avenida Trabalhador S\~{a}o-carlense, 400, S\~{a}o Carlos, 13566-590, Brazil, \printead{e1,e2}}

\end{aug}

\begin{abstract}
Bayesian Model Averaging (BMA) is an application of Bayesian inference to the problems of model selection, combined estimation and prediction that produces a straightforward model choice criteria and less risky predictions. However, the application of BMA is not always straightforward, leading to diverse assumptions and situational choices on its different aspects. Despite the widespread application of BMA in the literature,  there were not many accounts of these differences and trends besides a few landmark revisions in the late $1990$s and early $2000$s, therefore not taking into account any advancements made in the last $15$ years. In this work, we present an account of these developments through a careful content analysis of $587$ articles in BMA published between $1996$ and $2014$. We also develop a conceptual classification scheme to better describe this vast literature, understand its trends and future directions and provide guidance for the researcher interested in both the application and development of the methodology. The results of the classification scheme and content review are then used to discuss the present and future of the BMA literature.
\end{abstract}


\end{frontmatter}

\section{Introduction}


It is very common in practice that multiple models provide adequate descriptions of the distributions generating the observed data. It is standard statistical practice that, in such situations, a better model must be selected according to some criteria, like model fit to the observed dataset, predictive capabilities or likelihood penalizations such as information criteria. After selection is performed,  all inference is made and conclusions drawn assuming the selected model as the true model.

However, there are downsides to this approach. The selection of one particular model may lead to overconfident inferences and riskier decision making as it ignores the existent model uncertainty in favor of very particular distributions and assumptions on the model of choice. Therefore, modeling this source of uncertainty to appropriately select or combine multiple models is very desirable. 

Using Bayesian inference to this purpose has been suggested as a framework capable of achieving these goals \citep{Leamer1978}. Bayesian Model Averaging (BMA) is an extension of the usual Bayesian inference methods in which one does not only models parameter uncertainty through the prior distribution, but also model uncertainty obtaining posterior parameter and model posteriors using Bayes' theorem and therefore allowing for allow for direct model selection, combined estimation and prediction.

\subsection{Background}\label{S:Background}


Let each model in consideration be denoted by $M_l$, $l=1,\ldots,K$ representing a set of probability distributions encompassing the likelihood function  $L(\boldsymbol{Y}|\theta_l,M_l)$ of the observed data $\boldsymbol{Y}$ in terms of model specific parameters $\theta_l$ and a set of prior probability densities for said parameters, denoted in general terms by $\pi(\theta_l|M_l)$ on which we omit eventual prior hyperparameters for the sake of clarity. Notice that both the likelihood and priors are conditional on a particular model.

Given a model, one then obtains the posterior distribution using Bayes' theorem, resulting in 

\begin{equation}\label{E:posteriori1}
	\pi(\theta_l|\boldsymbol{Y},M_l) = \frac{L(\boldsymbol{Y}|\theta_l,M_l)\pi(\theta_l|M_l)}{\int_{ }L(\boldsymbol{Y}|\theta_l,M_l)\pi(\theta_l|M_l)d\theta_l}
\end{equation}

\noindent where the integral in the denominator is calculated over the support set for each prior distribution and represents the marginal distribution of the dataset over all parameter values specified in model $M_l$. 

This quantity is essential for BMA applications as we will show momentarily and is called the model's marginal likelihood or model evidence and is denoted by 

\begin{equation}\label{E:evidence}
\pi(\boldsymbol{Y}|M_l) = \int_{ }L(\boldsymbol{Y}|\theta_l,M_l)\pi(\theta_l|M_l)d\theta_l
\end{equation}

Bayesian model averaging then adds a layer to this hierarchical modeling present in Bayesian inference by assuming a prior distribution over the set of all considered models describing the prior uncertainty over each model's capability to accurately describe the data. If there is a probability mass function over all the models with values $\pi(M_l)$ for $l=1,\ldots,K$, then Bayes' theorem can be used to derive posterior model probabilities given the observed data by

\begin{equation}\label{E:posterior.P}
\pi(M_l|\boldsymbol{Y}) = \frac{\pi(\boldsymbol{Y}|M_l)\pi(M_l)}{\sum_{m=1}^{K}\pi(\boldsymbol{Y}|M_m)\pi(M_m)},
\end{equation}

\noindent resulting in a straightforward posterior model probability, representing the backing of each considered model by the observed data. 

There is also a link between these posterior model probabilities and the use of Bayes Factors. Given two models $l$ and $m$, the Bayes factor of model $l$ against model $m$ is given by

\begin{equation}
BF_{lm} = \frac{\pi(M_l|\boldsymbol{Y})}{\pi(M_m|\boldsymbol{Y})},
\end{equation}

\noindent thus quantifying the relative strength of the evidence in favor of model $l$ against that of model $m$. Given a baseline model, which we arbitrarily fix as model $1$, it is clear that equation \eqref{E:posterior.P} can be written in terms of Bayes Factors by simply dividing by the baseline model's evidence, resulting in 

\begin{equation}\label{E:posterior.BF}
\pi(M_l|\boldsymbol{Y}) = \frac{BF_{l1}\pi(M_l)}{\sum_{m=1}^{K}BF_{m1}\pi(M_m)},
\end{equation}

\noindent which means that one can estimate the posterior model probabilities by using estimates for Bayes Factors and vice versa.

These model probabilities can mainly be used for two purposes. First, the posterior probabilities \eqref{E:posterior.P} can be used as a straightforward model selection criteria, with the most likely model being selected. Second, consider a quantity of interest $\Delta$ present in all models, such as a covariate or future observation, it follows that its marginal posterior distribution across all models is given by

\begin{equation}\label{E:marginalests}
\pi(\Delta|\boldsymbol{Y}) = \sum_{l=1}^{K}\pi(\Delta|\boldsymbol{Y},M_l)\pi(M_l|\boldsymbol{Y}), 
\end{equation}

\noindent which is an average of all posterior distributions weighted by each posterior model probability. Therefore, BMA allows for a direct combination of models to obtain combined parameter estimates or predictions \citep{Roberts1965}. This practice leads to predictions with lower risk under a logarithmic scoring rule \citep{Madigan1994} than using a single model. 

However, the implementation and application of BMA is not without difficulties.  A prior distribution over the considered models must be specified, which is non trivial in most applications. 
Additionally, calculating each model evidence (equation \ref{E:evidence}) is non-trivial. Except in simple settings like in some generalized linear models with conjugate distributions, the evidence does not present a closed form and must be approximated, which presents plenty of challenges and is an active research field \citep{Friel2012}.

Despite these difficulties, BMA was extensively applied in the last $20$ years, mostly in combining multiple models for predictive purposes and selecting models, particularly covariate sets in regression models or network structure in Bayesian Network models. The latter application induces another pitfall in the form of large model spaces. For instance, consider a regression model with $p$ covariates. The number of possible models without any interaction coefficients is $2^p$, which represents a large number of models even for moderate values of $p$. This difficulty can be mostly addressed by prior filtering of all possible models or through stochastic search algorithms over the model space.

\subsection{Objectives of this review}\label{S:Obj}

As mentioned before, the idea of selecting and combining models based on their posterior probabilities is not news, but a series of advances made in the $1990s$ made the implementation and application of these ideas a reality. Following \cite{Leamer1978}, most model selection and marginal probabilities were only obtainable for the linear model under very specific parameter priors. However, the seminal work by \cite{Raftery1996} paved the way for a multitude of applications by providing a straightforward approximation for the evidence in generalized linear models. 

There were also advances in the implementation of BMA in large model spaces, from a preliminary filtering based on posterior probability ratios called Occam's Window \citep{Madigan1994} to a stochastic search algorithm inspired in the Reversible Chain Markov Chain Monte Carlo  \citep{Green1995} with trans-dimensional jumps based on posterior model probabilities (the MC$^3$ algorithm \cite{Madigan1995}). 
 

We also noticed that following the landmark reviews of \cite{Hoeting1999} and \cite{Wasserman2000}, there were no comprehensive reviews of the developments and applications of BMA in the last $15$ years, which does not account for the developments in Bayesian inference brought by the Markov Chain Monte Carlo revolution in the late $1990$s and $2000$s. With this paper, we aim to cover this undocumented period, specifically we have the following goals:

\begin{itemize}
\item To provide a conceptual classification scheme (CCS) to review and classify key components of the BMA literature
\item To summarize research findings and identify research trends
\item To obtain useful guidance for researchers interested in applying BMA to complex models
\end{itemize}

\subsection{Outline}

The remainder of this paper is structured as follows. In Section \ref{S:2}, we outline the literature search procedure and criteria employed to select the relevant BMA literature. The content of each selected article is then classified according to its main features using the Conceptual Classification Scheme described in Section \ref{S:CCS}, obtaining the patterns we describe and discuss in Section \ref{S:ResDis}. We conclude in Section \ref{S:Conc} with some guidance on the directions of the BMA literature. 

\section{Survey methodology}\label{S:2}



To better understand how BMA was applied, we chose to perform a content analysis of the published literature. The purpose of a content analysis is to systematically evaluate all forms of a recorded communication, identifying and classifying key contributions to a field and clarify trends, practice and indicate research possibilities. To achieve this objective, we formulated a systematic review following the guidelines in \cite{Moher2009} specifying objective criteria for defining the relevant literature to revise and appropriate ways to report our findings. 

\subsection{Literature search procedure}\label{S:LitSearch}

Aiming to perform a comprehensive search in the BMA literature, we combined four databases: Elsevier`s Scopus and ScienceDirect (available at \url{http://www.scopus.com} and \url{http://www.sciencedirect.com/} respectively), Thompsom Reuters`s Web of Science (available at \url{http://apps.webofknowledge.com}) and the American Mathematical Society`s MathSciNet database (available at \url{http://www.ams.org/mathscinet/index.html}). 

We performed queries  of the ``Bayesian Model Averaging" term restricted to the $1996-2014$ period on the publications' title, abstract and keywords for queries in Scopus and ScienceDirect, topic (encompassing title, abstract, keywords and proprietary ``keywords plus") in Web of Science and ``Anywhere" in MathSciNet, as it presented fewer search options. The time  period was chosen to cover most of the published literature not covered by previous works while still including seminal works.

Two exclusion criteria were employed over search results to select articles for further revision, namely:

\begin{enumerate}
\item Search results written in English, published in peer-reviewed journals as an article (which excludes conference proceedings, theses, dissertations, books, etc.) and available online.
\item Articles explicitly employing BMA, therefore excluding articles only listing BMA in keywords, alluding to BMA as a model combination procedure or applying BMA without further explanation or reference to the specific methodology employed. 
\end{enumerate}

\begin{figure}[hbtp]
\centering
\includegraphics[width=\linewidth]{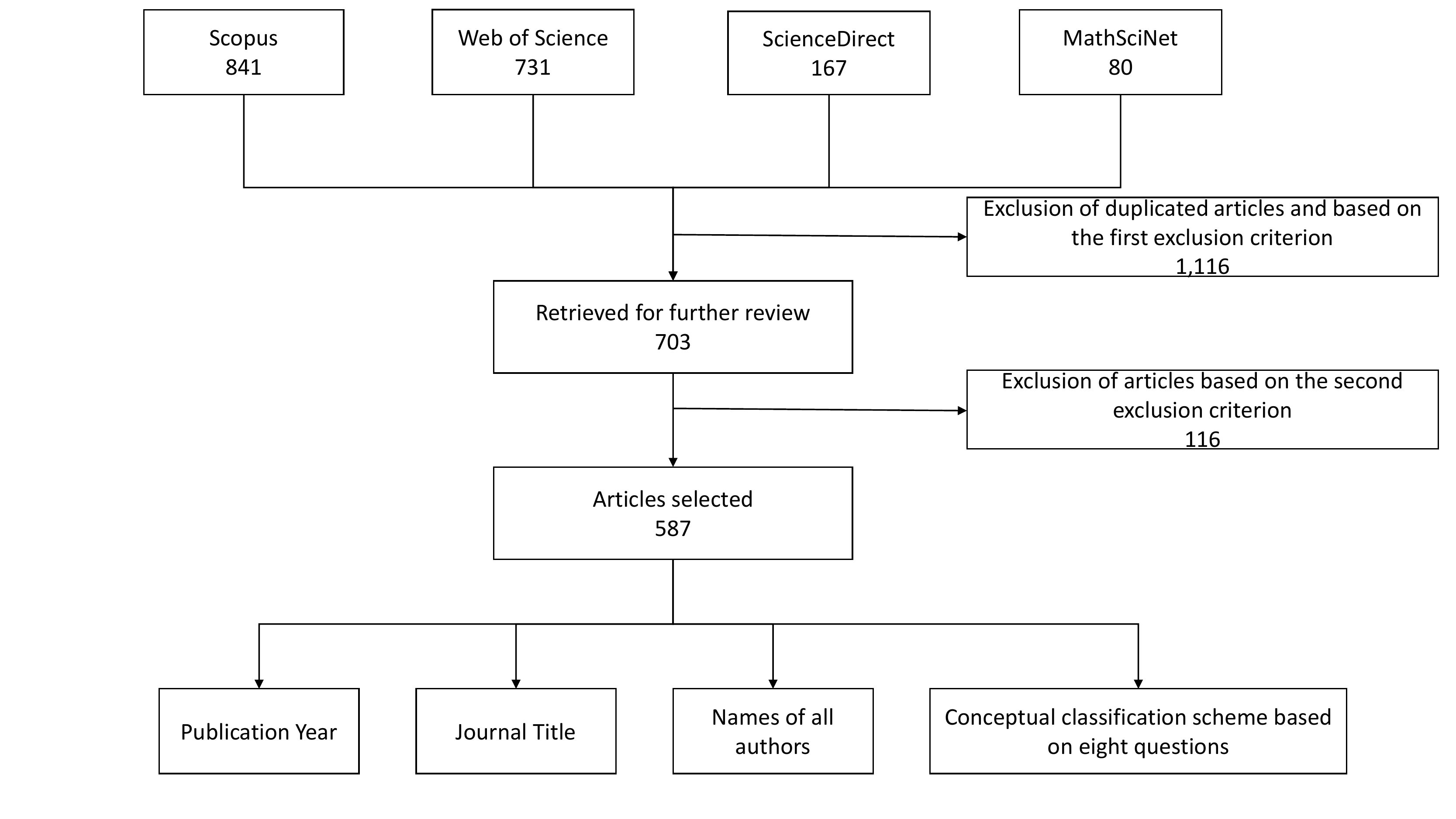}
\caption{Literature search procedure and number of selected articles}
\label{F:Fluxograma}
\end{figure}

Articles that did not comply with at least one of the criteria were excluded from the review.  The articles were then selected according to the procedure illustrated in Figure \ref{F:Fluxograma}. Using the aforementioned queries, $841$ articles were found in Scopus, $731$ in Web of Science, $167$ in ScienceDirect and $80$ in MathSciNet. After the removal of duplicates and application of the first exclusion criterion, $703$ articles were listed for further investigation  and carefully revised, leading to the exclusion of $116$ articles based on the second exclusion criterion, leaving $587$ articles for classification. 

\subsection{Structure of the dataset}

The final dataset consisting of $587$ eligible articles were then classified according to four main categories:
\begin{enumerate}
\item Publication year
\item Names of all authors
\item Journal title
\item Responses to the eight items of the Conceptual Classification Scheme (CCS)
\end{enumerate} 

\section{A conceptual scheme for BMA}\label{S:CCS}

Besides the systematic review of the literature, we also aimed to employ a content analysis of our finding, aiming to provide further inside into the applications of BMA, identify research trends and elucidate possible research directions and opportunities. To achieve this objective, we immersed ourselves in the selected dataset to better define characteristics in which current research that can be used to better understand the literature. This conventional content analysis \citep{Hsieh2005} allows us to better understand the literature without preconceived notions.

We codified these formulated characteristics into a conceptual classification scheme inspired in the scheme developed in \cite{Hachicha2012}, but adapted to characteristics relevant to the Bayesian Model Averaging literature. We then elaborated a Conceptual Classification Scheme (CCS) to classify the literature and refined the defined categories as we revised all articles. An article's classification according to the CCS is useful to a researcher interested in the field, as it gives key characteristics on the methodology and formulation, leading to more efficient queries into the developments and applications of BMA. In the present work, we elaborated a CCS with $8$ items which, along with the possible responses can be found in Table \ref{T:CCSquestions} and thoroughly described below.


\renewcommand{\labelenumii}{\theenumii}
\renewcommand{\theenumii}{\theenumi.\arabic{enumii}.}
\begin{table}[H]
\caption{List of questions and employed categories for the CCS \label{T:CCSquestions}}
\begin{tabular}{l}
\hline
\begin{minipage}[t]{0.8\columnwidth}%
\begin{enumerate}
\item How is BMA used? 
			\begin{enumerate}
				\item Model choice
				\item Combined parameter estimation
				\item Combined prediction 
				\item Conceptual discussion and methodological improvements
				\item Review of current methods
			\end{enumerate}
\item What is the field of application?
			\begin{enumerate}
				\item Statistics and Machine Learning
				\item Physical sciences and engineering
				\item Biological and medical sciences
				\item Economics and humanities
			\end{enumerate}
\item How are model prior specified? 
			\begin{enumerate}
				\item Vague prior
				\item Used verbatim from the literature
				\item Elicitated from experts or from the problem
				\item No explicit use or reference / not applicable
			\end{enumerate}
\item How is the evidence estimated? 
			\begin{enumerate}
				\item Monte Carlo sampling and extensions
				\item Analytical approximations 
				\item Markov Chain Monte Carlo 
				\item Ratio of densities  
			\end{enumerate}
\item How do the authors real with high dimensionality? 
			\begin{enumerate}
				\item Dimensionality reduction 
				\item Stochastic search via Markov Chain Monte Carlo 
			\end{enumerate}
\item Are Markov Chain Monte Carlo based methods employed?
			\begin{enumerate}
				\item Yes
				\item No
			\end{enumerate}
\item Is the method validated through simulation studies? 
			\begin{enumerate}
				\item Yes
				\item No
			\end{enumerate}
\item Is the application validated through data-driven procedures?
			\begin{enumerate}
				\item Cross validation and data splitting
				\item Leave-one-out or K-fold cross validation
				\item Posterior predictive checks 
				\item None / not applicable
			\end{enumerate}
\end{enumerate} %
\end{minipage}\tabularnewline
\end{tabular}
\end{table}

\subsection{Usage}

Being  a framework for model selection and combination, BMA can be used to solve a wide range of problems. As such, we first classified each eligible article with respect to usage. Our content search resulted in five main categories of BMA usage. 

The derivation of posterior model probabilities make BMA a very straightforward model choice method. Said probabilities are easily interpretable under the Bayesian paradigm and the model space can be as wide as necessary and it requires no bookkeeping over the number of parameters or kind of penalty as required by the information criteria methods applied in the statistical literature. We therefore singled out articles that employ BMA as a model choice method. 

Using the posterior model probabilities as weights, one can produce estimates averaged over all considered models (equation \eqref{E:marginalests}), that can have a lower overall risk and take model uncertainty into consideration. There are two main kind of estimates. When $\Delta$ in equation \eqref{E:marginalests} represents a parameter that is common to all considered models, one can obtain an averaged model estimate. On the other hand, when $\Delta$ represents a future observation or yet to be collected data point, equation \eqref{E:marginalests} can be used to derive an average prediction. Albeit similar, both applications have distinct uses so we classified articles on its use of BMA to obtain joint estimation and prediction, respectively.

As we previously mentioned, BMA is not as direct to apply. As such, there is a considerable technical literature in theoretical aspects and extensions of BMA. There is also retrospective studies in which BMA or any one of the applications mentioned above are revised in a specific area (model selection methods in economics, for instance) so we classified these articles as conceptual discussions and  review articles respectively.

\subsection{Field of application}\label{SS:Field}

Given the wide array of possible uses, we also classified every selected article according to its field of application. We employed a vague classification of BMA applications for two reasons. First, as stated in Section \ref{S:Obj}, we aim to summarize and point research trends, not perform a detailed taxonomy of BMA usage. Second, as BMA (and Bayesian methods in general) find more penetration into applied research, we lack the expertise to discriminate between subfields. Restricting the present work to a more general view, we defined four fields of application.

An immediate field of application for BMA is the Statistics literature, which is in turn heavily borrowed and borrows upon the Machine Learning literature. Therefore, we classified papers more concerned with statistical modeling, theoretical developments and machine learning applications into a single category. After some pioneering works in the statistical literature, BMA found its way into the Biological and Life Sciences, comprising of the studies in the fields of Medicine, Epidemiology, Biological sciences such as Ecology and others. The seminal revision work by \cite{Geweke1999} introduced BMA to the field of Economics and later on to other Humanities such as Political and Social sciences. After a few years, some works started to appear in the fields of Engineering and Physical sciences such as Meteorology and Physics. 


\subsection{Model priors}

Much of the study in Bayesian methods is concerned with the elicitation of prior distributions, representing the uncertainty in place before any data is observed. Elicitating an appropriate prior is a non-trivial task in any Bayesian setting, and such difficulties are compounded in model averaging since a probability measure for the model space is not obvious in principle, which is probably reflected in the reluctance by many authors to explicitly state their priors. Throughout our review, we encountered four main categories.

One can simply assume prior ignorance about which model is correct through a vague prior (i.e., $\pi(M_l) \propto 1$, $l=1,2,\ldots,K$), that assumes no model is more likely  \textit{a priori} than any other and let the observed data carry all the information. This is not always desirable and sometimes it is possible to perform some elicitation based on expert opinions or specific characteristics of a particular problem, resulting in elicitated priors. 

In the case of elicitated, conjugate or simply convenient priors, sometimes a particular choice spreads through the literature, resulting in subsequent authors that use such priors verbatim, which we classified into a single category of a literature prior. Finally, as mentioned previously a number of authors bypass the problem entirely by neither explicitly stating their model priors nor providing references to clarify their assumptions yet still use BMA in some sense. We classified these cases as ``not available" (NA).  

\subsection{Evidence estimation} \label{S:Evidence}

The estimation of the marginal likelihood (equation \eqref{E:evidence}), also known as the model evidence is non-trivial in any general setting since  it usually involves complicated  multidimensional integrands with sometimes complex support sets that can make the integration unfeasible. As such, many solutions have been proposed, which we classified into five categories.

The integral can be approximated by Monte Carlo methods or extensions such as Importance Sampling. 
Namely, given an importance probability density $w(\theta)$ defined over the same integration domain of equation \eqref{E:evidence}, the evidence can be approximated by taking $R$ random samples $\theta_1,\ldots, \theta_{R}$ from the probability distribution determined by $w(\theta)$ and computing the weighted average

\begin{equation}
\widehat{\pi(\boldsymbol{Y}|M_l) } = \frac{1}{R}\sum^{R}_{r=1}\frac{L(\boldsymbol{Y}|\theta_r,M_l)\pi(\theta_r|M_l)}{w(\theta_r)}
\end{equation}

\noindent which is guaranteed to converge to the evidence by the Strong Law of Large Numbers.  Clearly, the ordinary Monte Carlo approximation can be performed by using the parameter prior as an importance density. Further extensions of this idea exist in the form of Bridge sampling \citep{Gelman1998} and other forms of integration. Relying on the same Monte Carlo framework, we classified these methods into one category,

Still in the spirit of Monte Carlo methods but very distinct in practice, one could also sample from a Markov Chain, using Markov Chain Monte Carlo (MCMC) methods to approximate the evidence. Let $\theta^{(1)}_l,\ldots,\theta^{(R)}_l$ be $R$ posterior samples obtained from a MCMC chain and $w(\theta)$ an importance density as defined previously, then one can use the ``Importance Sampling'' estimator \citep{Gelfand1994}

\begin{equation}\label{E:GelfandDey}
\widehat{\pi(\boldsymbol{Y}|M_l) } = \left\{\frac{1}{R}	\sum_{r=1}^{R}\frac{w(\theta^{(r)}_l)}{L(\boldsymbol{Y}|\theta^{(r)}_l,M_l)\pi(\theta^{(r)}_l|M_l)}	\right\}^{-1},
\end{equation}

\noindent which is, in turn, a generalization of the harmonic mean estimator of \cite{Newton1994} that uses the prior as an importance density. The estimator is shown to converge to the evidence as the sample size increases, but an importance function must be finely tuned to avoid estimators with unbounded variance. 

Another flavor of MCMC posterior probability estimator comes through the use of trans-dimensional Markov Chain methods like the Reversible Jump MCMC \citep{Green1995} or stochastic searches through the model space like variable selection  through stochastic search (SSVS,\cite{George1993,George1997}) employed in regression models. 

In these methods, multiple models are sampled  either through a Gibbs Sampler or a Metropolis jump on the same MCMC chain. Consider the output of a MCMC procedure of $R$ posterior samples and let $\gamma_r$ be a variable indicating which model the chain is visiting at step $r$, $\gamma_r \in \left\{1,\ldots,K\right\}$ . The model posterior is estimated commonly by the sample mean 

\begin{equation}
\widehat{\pi(\boldsymbol{Y}|M_l)} = \frac{1}{R}  \sum_{r=1}^{R}I\left(\gamma_r = l\right)
\end{equation}

\noindent where $I(\cdot)$ is the indicator function or by some Rao-Blackwellized estimator when available \citep{Guan2011}. 
When using MCMC samples, the quality of the approximation is not guaranteed and there are more sophisticated results ensuring its good behavior (see \cite{Robert2013} for a complete treatment). In this paper, we classified all MCMC based methods into a single category.

Also of particular interest is the ratio of densities method popularized by \cite{Chib1995}, in which one exploits the fact that equation \eqref{E:posteriori1} is valid for every parameter value, whereas the normalizing constant stays the same. \cite{Chib1995} then suggests picking one particular parameter point $\theta^{*}$ and estimate the evidence as 

\begin{equation}
\widehat{\pi(\boldsymbol{Y}|M_l) } = \frac{L(\boldsymbol{Y}|\theta^{*}_l,M_l)\pi(\theta^{*}_l|M_l)}{\pi(\theta^{*}_l|\boldsymbol{Y},M_l)},
\end{equation}

\noindent for $l=1,\ldots,K$ and the parameter value  $\theta^{*}$ is chosen as to minimize estimation error.

Besides stochastic approximations, analytical approximations based on asymptotic results can be employed  as shown in the seminal works by \cite{Kass1995} and \cite{Raftery1996}. Based on a Taylor expansion around the posterior mode and imposing some regularity conditions, one can approximate the evidence through the Laplace approximation. Namely, if $\tilde{\theta}_l$ is the posterior mode for model $l$, then 

\begin{equation}
\pi(\boldsymbol{Y}|M_l) \approx (2\pi)^{\frac{p_l}{2}}\sqrt{|\Psi_l|}L(\boldsymbol{Y}|\tilde{\theta}_l,M_l)\pi(\tilde{\theta}_l|M_l),
\end{equation}

\noindent where $p_k$ is the number of parameters in model $l$ and $\Psi_l$ is minus the inverse Hessian matrix of the log-posterior given by $\log\left(L(\boldsymbol{Y}|\tilde{\theta}_l,M_l)\pi(\tilde{\theta}_l|M_l)\right)$ for the model. Under regularity conditions, the approximation is $O(n^{-1})$. Let  $\hat{\theta}_l$ denote the maximum likelihood estimator for model $l$, then the Bayes Factor between two models $l$ and $m$ can be reasonably approximated by the Bayesian information criteria (BIC), given by

\begin{equation}
2\log B_{lm} \approx 2\left(\log\left(L(\boldsymbol{Y}|\hat{\theta}_l,M_l) \right) - \log\left(L(\boldsymbol{Y}|\hat{\theta}_m,M_m)\right)\right) - (p_l - p_m) \log N
\end{equation}

\noindent when both models are used to fit the same dataset of sample size $N$. The BIC provides a good approximation for many generalized linear models and enjoys widespread use, even with the larger approximation error of $O(1)$.  Both methods are very similar in spirit, and as such, were put into the same classification.

Finally, many authors were able to compute the evidence in closed form, either by the shape or their models and distributional assumptions (as in the linear model) or through the use of convenient parameter prior distributions. Since we were more interested in the approximation of complex evidences in general settings, we classified these cases as non-applicable (NA).

\subsection{Dimensionality}\label{S:Dim}

It is very common in model selection settings to encounter extremely wide model spaces in which the exhaustive fit of all models is unfeasible. We illustrate the problem with one of the most popular applications of model choice, variable selection in regression models. Let $Y$ represent an observation and $\boldsymbol{X}$ represent a $p \times 1$ vector of covariates which we aim to investigate the degree of association to $Y$ through the linear model

\begin{equation}
Y = \beta \boldsymbol{X} + e,
\end{equation}

\noindent where $\beta$ is a $1 \times p$ parameter vector of fixed effects and $e$ is a random residual effect. It follows that finding the subset of covariates (most) associated with $Y$ induces a model selection problem. 

However, not considering any interaction terms, the number of possible subsets (and  therefore models) is $2^p$, which grows geometrically with the number of covariates resulting in very large model spaces even for a moderate number of covariates precluding an exhaustive investigation of all models. The BMA literature answered to this problem through two main approaches: dimensionality reductions of the model space and stochastic searches through MCMC. 

One of the first dimensionality reduction techniques was the Leaps and Bounds algorithm \citep{Furnival1974}. The algorithm aims to select the best performing regression models based on the residual sum of squares, with the model with smaller sum being selected as more fit to the data. Exploiting relationships on the linear model and the Branch and Bound optimization algorithm, the most parsimonious models can be obtained without an exhaustive search. In much of the literature, a preliminary search through the Leaps and Bounds algorithm is performed to subject only the most promising models to BMA.

Another popular criteria is the Occam's Window \citep{Madigan1994}. This criteria argues that only models with a relatively high posterior probability must be considered. As such, it reduces the model set comprised by the $K$ models to the reduced set

\begin{equation}
A = \left\{M_k : \frac{\max_{l}P(M_l|\boldsymbol{Y})}{P(M_k|\boldsymbol{Y}) }\leq c\right\},
\end{equation}

\noindent where $c$ is a tuning parameter chosen by the user. The Occam's window excludes all models whose model probability is smaller than the most likely model by a factor of $c$, usually set to $20$ to emulate the popular $0.05$ cutoff employed when filtering models based on $p-$values. Note that, albeit straightforward to set up, the application of Occam's Window relies on the easy calculation or approximation of the model posterior probabilities, which in turn relies on the model evidence. In our review, we classified all aforementioned preliminary filtering methods together.

Instead of reducing the model space beforehand, other approaches take the entire model head on and perform some kind of search, mostly through MCMC techniques. In these cases, not only the dimensionality problem is dealt with, but there is also a straightforward manner to estimate the posterior model probabilities through the proportion of times a model is visited throughout the search.

One of the first proposals for these searches in regression models was proposed by \cite{George1993} called Stochastic Search Variable Selection (SSVS). The SSVS utilizes a set of auxiliary random variables $\gamma_l$, $l=1,\ldots,p$ such that

\begin{equation}
\gamma_l = \left\{\begin{array}{l}
1,\text{ if variable $l$ belongs in the model}\\
0, \text{ otherwise}
\end{array}\right. .
\end{equation}

Along with all other parameters, a prior distribution is assigned to these variables and therefore a posterior distribution is obtained through MCMC procedures. The stochastic search is performed by the updating of the indicator variables, each configuration representing a distinct model. SSVS just expands over an already implemented MCMC algorithm, making it a widely used and flexible methodology.

A more general approach was proposed by \cite{Green1995} in the Reversible Jump MCMC (RJMCMC) algorithm, which just like in the current BMA setting, insert all models into a larger space in which a Markov Chain is constructed. Model search is then performed using two components: a proposal probability density for a model $l$ given the current model $M_m$  and an invertible differentiable mapping between the two spaces defined by the models. The chain then moves through models by means of a Metropolis-Hastings step. The construction of the acceptance probability required for the MH step is not straightforward and we shall omit it for the sake of clarity. The interested reader is directed to \cite{Green1995} for a complete treatment.

There is, however, a more straightforward and relatively general model search procedure in the literature. The Markov Chain Monte Carlo Model Composition (MC3,\cite{Madigan1995}) in which one applies the same idea of a Metropolis-Hastings step for model jumps from RJMCMC but in a simplified fashion. Let $M_l$ and $M_m$ be two models, MC3 performs a model change from model $M_l$ to model $M_m$ with acceptance probability

\begin{equation}
\alpha(m,l) = \min \left\{1,\frac{P(M_m|\boldsymbol{Y})}{P(M_l|\boldsymbol{Y})}\right\}.
\end{equation}

 Given that marginal probabilities are available, the implementation of MC3 is straightforward. For the purposes of our categorization scheme, we considered all methods employing MCMC based searches similar and therefore were classified into a single category.

Throughout our literature review, we also encountered applications in which dimensionality was not an issue. In these cases, all models were fit to the data and BMA performed posteriorly. As these articles obviously did not propose any way to mitigate dimensionality problems, we classified these articles as ``not applicable" (NA).

\subsection{Markov Chain Monte Carlo methods}

It is impossible to ignore the revolution in Bayesian Inference sparked by the dissemination of MCMC methods and software. MCMC made it possible to perform inference with very complex likelihood and prior structures and obtain estimates of key posterior quantities based on straightforward outputs and grounded on solid theory, making it the default approach for many applied problems. 

Such popularity comes in part from the popularization of out-of-the-box MCMC software containing robust MCMC implementations to a wide range of problems like the widely used ``Bayesian inference Using the Gibbs Sampler" software (BUGS, \cite{Spiegelhalter1995}) and the ``Just Another Gibbs Sampler" (JAGS, \cite{Plummer2003}). These software, commonly integrated with the R Statistical Software, made MCMC methods and Bayesian inference available to a broad audience. We aimed to track this spread of MCMC methods through the BMA literature by classifying each article on its usage of MCMC methods.

\subsection{Simulation studies}

We classified articles on the practice of generating simulations from the proposed models or the use of artificial datasets. Simulation studies can be employed to investigate characteristics of the averaging process and desired properties like predictive power in the best case scenario, emulate physical systems to better understanding and generate predictions from diverse models for averaging. As such, we classified articles with respect to the presence of simulation studies.

\subsection{Data-driven validation}

After BMA was applied, we also investigated how the process was validated using real data. The most traditional data-driven validation procedure consists of simply splitting the dataset into at least two disjoint sets, fitting the model to the data on the former and validating the fitted model on the latter. This kind of cross validation is very commonly used and articles practicing this kind of validation were put into a single category. 

Another category was made for more sophisticated kinds of cross validation like $K-$fold cross validation. In this procedure, the dataset is split into $K$ disjoint subsets and, for each subset, the model is fit over the combined remainder $K-1$ subsets. The chosen subset is then used as a validation set and the process is repeated in turn for the next subset. After going through all subsets, the validation measures employed (goodness-of-fit, predictive power, etc) are somehow combined for an overall conclusion. Being significantly more complex than simple data splitting, we classified articles using $K-$fold cross validation or special cases like the popular leave-one-out cross validation into the same category.

Being a Bayesian procedure in nature, it is not unexpected that applications of BMA might use Bayesian goodness-of-fit measures like Posterior Predictive Checks \citep{Gelman1996}. For the predictive check, one chooses a test statistic over the observed dataset and compares it with replicated data generated using the posterior distribution and a model is said do present good fit if, averaged over the posterior, the test statistic is not too extreme when compared to its value calculated over the observations. One usually generates the replicated data using posterior samples obtained from MCMC methods so that  required averages are straightforward from the estimation procedure. This procedure is clearly distinct from the previous validation procedures mentioned and, therefore, categorized separately.

\section{Results and discussion}\label{S:ResDis}

We first performed some descriptive statistics to investigate the growth and some trends in the BMA literature and then investigated the patterns in the light of our proposed CCS.

\subsection{Descriptive statistics for the BMA literature}

\begin{figure}[hbtp]
\centering
\includegraphics[width=\linewidth]{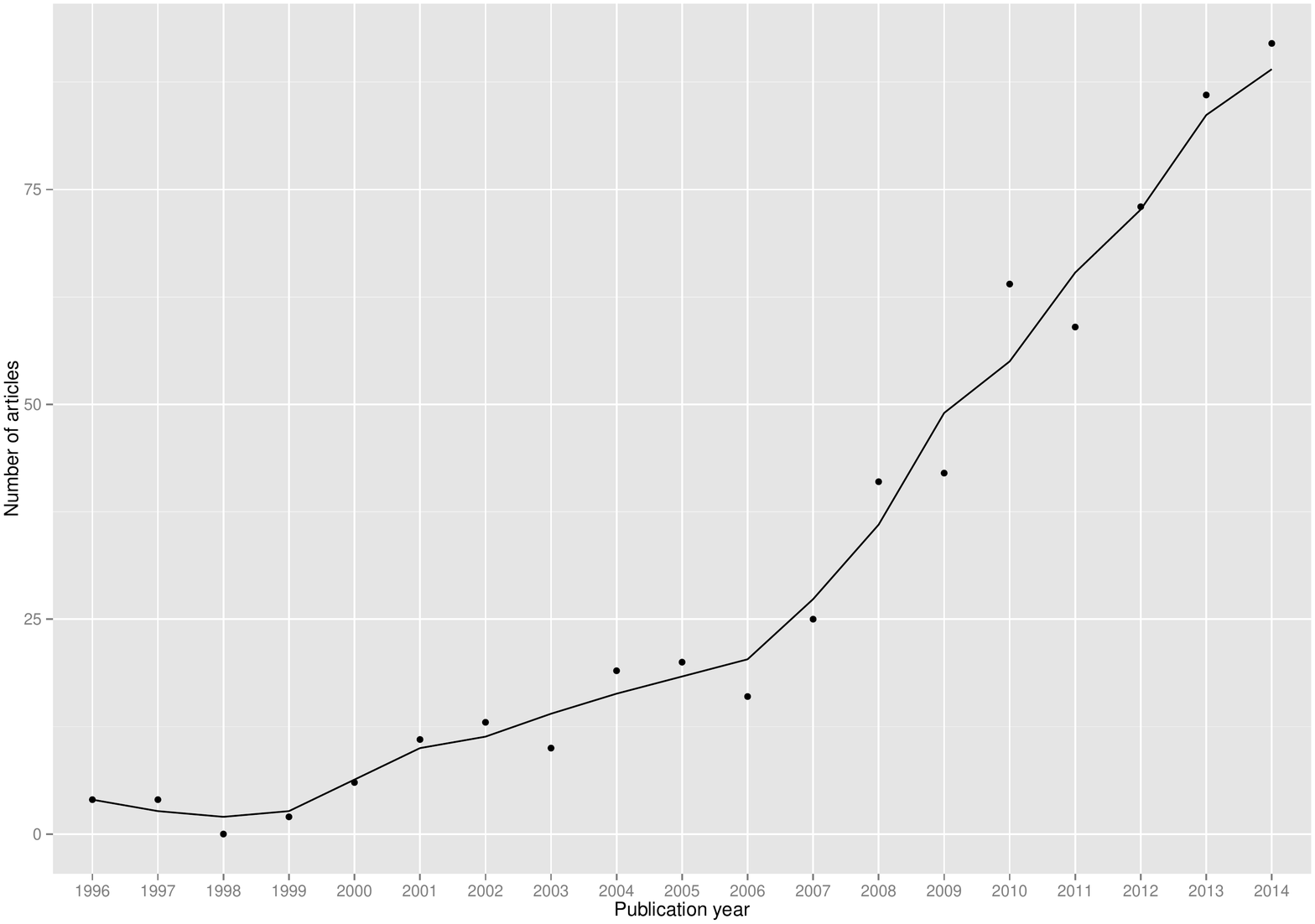}
\caption{Number of published articles by year of publication and $3$-year moving average}
\label{F:Numero}
\end{figure}

We segmented our dataset with respect to publication year to investigate the growth of the literature in terms of the number of published articles throughout the considered period. Since peer-reviewed articles can take a significant time from submission to publication, we smoothed the temporal trends using a $3$ year moving average. The results can be observed in Figure \ref{F:Numero}.

One can interpret the growth shown in figure \ref{F:Numero} in three stages. First, in the $1996-2000$ time period, there were a relatively small number of publications, not due to the lack of theoretical results, but rather for the absence of a systematic exposition to BMA and accessible computational tools. Then, with the publication of the revisions by \cite{Hoeting1999}, \cite{Wasserman2000} and \cite{Geweke1999} and some computational tools implementing the results by \cite{Raftery1996}, \cite{Volinsky1997}, \cite{Madigan1994} and \cite{Madigan1995} led to a popularization of BMA resulting in a growth in publications in the $2000-2005$ time period. Then, after $2005$, there was a veritable increase in the number of publications, probably due to the widespread use of the aforementioned computational tools, more easily accessible computer power and the increasing availability of ready-to-use MCMC software.


We then divided the database according to the journal title, to identify fields with a more widespread application of BMA. There was no single periodical responsible for most of the literature and it is clear that its applications have spread through diverse fields. The $100$ titles with the higher counts of publications are listed in Table \ref*{T:Periodicals}.

There are some clusters representing some patterns in the diffusion and application of BMA. As it is expected from a statistical methodology, there were many articles published in Statistics and Machine Learning periodicals, adding up to $21$ periodicals in the top $100$. There is also a widespread application of BMA in the Economics literature, with  $21$ titles in the top $100$ and a few periodicals in Meteorology and Climatology, following the seminal predictive framework proposed by \cite{Raftery2005}.

\begin{longtable}{lll}
\caption{Number of publications on the $100$ periodicals with most BMA articles \label{T:Periodicals}}  \\
\hline
 Title & Articles published & Percentage \\ 
  \hline
 Weather Resources Research & 18 & 3.07 \\ 
 Monthly Weather Research & 15 & 2.56 \\ 
 Journal of the American Statistical Association & 12 & 2.04 \\ 
 Journal of Applied Economics & 11 & 1.87 \\ 
  Computational Statistics and Data Analysis & 10 & 1.7 \\ 
  Journal of Hydrology & 10 & 1.7 \\ 
  Biometrics & 9 & 1.53 \\ 
  Economic Modeling& 8 & 1.36 \\ 
  Canadian Journal of Fisheries and Aquatic Sciences & 7 & 1.19 \\ 
  Journal of Econometrics & 7 & 1.19 \\ 
  Statistics in Medicine & 7 & 1.19 \\ 
  Biometrika & 6 & 1.02 \\ 
  Journal of Applied Statistics & 6 & 1.02 \\ 
  Ecological Economics & 6 & 1.02 \\ 
  Journal of Forecasting & 6 & 1.02 \\ 
  Journal of Hydrologic Engineering & 6 & 1.02 \\ 
  Neuroimage & 6 & 1.02 \\ 
  Public Library of Science - One (PLOS-ONE) & 6 & 1.02 \\ 
  Advances in Water Research & 5 & 0.85 \\ 
  International Journal of Forecasting & 5 & 0.85 \\ 
  Journal of Agricultural Biological and Environmental Statistics & 5 & 0.85 \\ 
  Journal of International Money and Finance & 5 & 0.85 \\ 
  Journal of Macroeconomics & 5 & 0.85 \\ 
  Risk Analysis & 5 & 0.85 \\ 
  Stochastic Environmental Research and Risk Analysis & 5 & 0.85 \\ 
  Annals of Applied Statistics & 4 & 0.68 \\ 
  Applied Economics & 4 & 0.68 \\ 
  Bioinformatics & 4 & 0.68 \\ 
  Ecological Applications & 4 & 0.68 \\ 
  Environmental Modeling and Software & 4 & 0.68 \\ 
  Genetic Epidemiology & 4 & 0.68 \\ 
  Hydrology and Earth System Sciences & 4 & 0.68 \\ 
  Journal of Computational and Graphical Statistics & 4 & 0.68 \\ 
  Journal of Machine Learning and Research & 4 & 0.68 \\ 
  Journal of the Royal Statistical Society-B & 4 & 0.68 \\ 
  Technometrics & 4 & 0.68 \\ 
  Applied Economics Letters & 3 & 0.51 \\ 
  BMC-Bioinformatics & 3 & 0.51 \\ 
  Conservation Biology & 3 & 0.51 \\ 
  Cerebral Cortex & 3 & 0.51 \\ 
  Climate Dynamics & 3 & 0.51 \\ 
  Economic Journal & 3 & 0.51 \\ 
  Global and Planetary Change& 3 & 0.51 \\ 
  Journal of Applied Metereology and Climatology & 3 & 0.51 \\  
  Journal of Business and Economic Statistics & 3 & 0.51 \\ 
  Journal of Banking and Finance & 3 & 0.51 \\ 
  Journal of Gerontology-A & 3 & 0.51 \\ 
  Journal of Geophysical Research: Atmospheres & 3 & 0.51 \\ 
  Machine Learning & 3 & 0.51 \\ 
  Quarterly Journal of the Royal Metereological Society & 3 & 0.51 \\ 
  The Review of Economics and Statistics & 3 & 0.51 \\ 
  Statistics and Computing & 3 & 0.51 \\ 
  Statistical Science & 3 & 0.51 \\ 
  Tellus-A & 3 & 0.51 \\ 
  Annals of Human Genetics & 2 & 0.34 \\ 
  Asia and Pacific Journal of Atmospheric Sciences & 2 & 0.34 \\ 
  Applied Stochastic Models in Business and Industry & 2 & 0.34 \\ 
  Bayesian Analysis & 2 & 0.34 \\ 
  BMC-Systems Biology & 2 & 0.34 \\ 
  Biostatistics & 2 & 0.34 \\ 
  Canadian Journal of Fisheries & 2 & 0.34 \\ 
  Clinical Trials & 2 & 0.34 \\ 
  Economics Bulletin & 2 & 0.34 \\ 
  Environmetrics & 2 & 0.34 \\ 
  Environmental Monitoring and Assessment & 2 & 0.34 \\ 
  Econometric Reviews & 2 & 0.34 \\ 
  Economic Systems & 2 & 0.34 \\ 
  Ensayos sobre politica economica & 2 & 0.34 \\ 
  Freshwater Biology & 2 & 0.34 \\ 
  Forest Ecology and Management & 2 & 0.34 \\ 
  Fisheries Research & 2 & 0.34 \\ 
  Geophysical Research Letters & 2 & 0.34 \\ 
  Genetics Selection Evolution & 2 & 0.34 \\ 
  Human Heredity & 2 & 0.34 \\ 
  Hydrological Processes & 2 & 0.34 \\ 
  Institute of Electrical and Electronics Engineers-B & 2 & 0.34 \\ 
  Institute of Electrical Engineers Transactions & 2 & 0.34 \\ 
  International Journal of Distributed Sensor Networks & 2 & 0.34 \\ 
  International Journal of Environmental Research and Public Health & 2 & 0.34 \\ 
  Journal of Animal Science & 2 & 0.34 \\ 
  Journal of Biogeography & 2 & 0.34 \\ 
  Journal of Computational Physics & 2 & 0.34 \\ 
  Journal of Environmental Economics and Management & 2 & 0.34 \\ 
  Journal of Economic Growth & 2 & 0.34 \\ 
  Journal of Economic Surveys & 2 & 0.34 \\ 
  Journal of the Japanese Statistical Society & 2 & 0.34 \\ 
  Journal of Money, Credit and Banking& 2 & 0.34 \\ 
  Journal of the Royal Statistical Society-A & 2 & 0.34 \\ 
  Journal of the Royal Statistical Society-C & 2 & 0.34 \\ 
  Journal of the Royal Statistical Society-D & 2 & 0.34 \\ 
  Marine Mammal Science & 2 & 0.34 \\ 
  Monthly Notices of the Royal Astronomical Society & 2 & 0.34 \\ 
  Political Analysis & 2 & 0.34 \\ 
  Public Library of Science- Computational Biology & 2 & 0.34 \\ 
  Physical Review D & 2 & 0.34 \\ 
  The Review of Financial Studies & 2 & 0.34 \\ 
  Regional Studies & 2 & 0.34 \\ 
  Systematic Biology & 2 & 0.34 \\ 
  Others & 210 & 35.78 \\ \hline
  Total & 587 & 100 \\ 
   
\end{longtable}

We also classified each article with respect to the authors and co-authors and listed the top $10$ most productive authors in Table \ref{T:Auteurs} to infer the most massive contributions to the literature. There were no authors overwhelmingly present in the literature given its size, but one can single out the contributions of Adrian Raftery from the University of Washington. Raftery authored or co-authored $34$ articles in our dataset, more than three times the second most productive author and contributed to the theoretical underpinnings of BMA \citep{Kass1995,Raftery1996} and to its applications, for instance in genetics \citep{Yeung2005}, engineering \citep{Raftery2010}, economics \citep{Eicher2011} and proposed an ensemble prediction method in meteorology and climatology \citep{Raftery2005} that enjoys widespread use.

\begin{table}[h]
\caption{Top $10$ authors in the BMA literature in number of publications\label{T:Auteurs}}
\centering
\begin{tabular}{lll}
  \hline
 Name & Institution & Number of authored/co-authored articles \\ 
  \hline
Adrian E. Raftery & University of Washington & 34 \\ 
Tilmann Gneiting & Karlsruhe Institute of Technology  & 11 \\ 
Merlyse Clyde & Duke University & 10 \\ 
Q. J. Wang & CSIRO Land and Water & 9 \\ 
Theo S. Eicher & University of Washington & 8 \\ 
Gary Koop & University of Strathclyde & 8 \\ 
 David Madigan & Columbia University & 7 \\ 
Xuesong Zhang & Pacific Northwest National Laboratory & 7 \\ 
 Edward L. Boone &  Virginia Commonwealth University & 6 \\ 
 Jesus Crespo-Cuaresma & Vienna University of Economics and Business& 6 
\end{tabular}
\end{table}

Following, there are Tilmann Gneiting who mainly contributed to ensemble methods in Meteorology, Merlyse Clyde, who applied BMA mostly in the context of variable selection for regression models using Stochastic Search Variable Selection, Q. J. Wang with contributions in Metereology, Theo S. Eicher  and Gary Koop , who applied BMA to Economics, David Madigan, who was part of the seminal BMA works in the late $1990s$, Xuesong Zhang  with contributions in Hydrology, Edward L. Boone who used BMA in Ecology and Jesus Crespo-Cuaresma in Economics.

\subsection{Conceptual classification scheme}

We present below some brief descriptive statistics and discussion on the classification patterns generated using the Conceptual Classification Scheme (CCS) to the dataset, with some notes to trends and guidance to specific applications of BMA. It is not useful to list the hundreds of revised articles, so we aim to provide illustrative works for each aspect of our revision as a the interested reader can follow through its references and citations for his own purposes. Furthermore, the dataset of all responses to the CCS can be obtained upon request from the corresponding author and full citations of the revised articles can be found as a Supplement to this article. 

\subsubsection{Usage}

The most common usage of BMA in the revised literature was model choice, with $231$ works totaling almost $40\%$ of all articles and the overwhelming majority of the revised articles deal with model choice through variable selection in regression models. Overall, we could spot three overarching themes in model choice throughout the literature with eventual variations. 

Firstly, there is the application of the background introduced by Adrian Raftery and collaborators in the first half of the $1990s$. These works perform dimensionality reduction of the model space through Occam's Window or Leaps and Bounds and approximate the model evidence using BIC like \cite{Volinsky1997}, in which variable selection and model averaging is performed for a Cox regression model. This set of techniques enjoy great popularity to this date due in part to its implementation in the \texttt{BMA} R package \citep{Raftery2005b}.Secondly, there are many works concerned with model selection in the linear model, specially in economical applications. Either all $2^p$ possible models are considered or there is a stochastic search using MC3 and model evidences are derived explicitly from a conjugate priors \citep{Fernandez2001b}. Finally, there is model choice using stochastic search through MCMC methods, like Reversible Jump MCMC over spaces with different numbers of covariates \citep{Lunn2008} and Stochastic Search Variable Selection \citep{Brown2002}.


After model choice, the most popular usage was the combination of multiple models for  prediction, which was performed in $161$ articles (around $28\%$ of the dataset). While combining each model's prediction is straightforward in principle, we identified at least three different trends. 

First, there is the straightforward application of BMA by fitting all models to the data, calculating model evidences, generating a prediction from each model's predictive distribution culminating in a combined prediction. This practice leads to lower risk predictions under a logarithmic loss \citep{Madigan1994} and is relatively widespread in the literature, with applications in Ecology \citep{Wintle2003} and Genetics \citep{Annest2009}. Secondly, there is a compromise between variable selection and prediction through an application of BMA to select the models or covariates with highest posterior model probability and the selected model is then used to derive predictions. This procedure employs the machinery developed for variable selection in favor of prediction, like SSVS \citep{Lamon2000} or RJMCMC \citep{Jacobson2004}. Finally, there is an alternative use of BMA ideas proposed in \cite{Raftery2005} for meteorological applications that differs somewhat of usual applications and that we will briefly discuss below.

The authors consider the problem of combining forecasts from $K$ models, $f_1,\ldots,f_K$ into one combined prediction. For each forecast there is a probability density function for the quantity one wished to forecast $y$ denoted by $g_l(y|f_l)$ for $l=1,\ldots,K$. \cite{Raftery2005} then proposes to construct a combined density function using the weighted average 

\begin{equation}
g(y|f_1,\ldots,f_K)  = \sum_{l=1}^{K}w_l g_l(y|f_l),
\end{equation}

\noindent in which $\sum_{l=1}^{K}w_l =1$, and the weights are interpreted in an analogous fashion to the posterior model probabilities in usual BMA. Assuming then each density as a normal, the authors estimate the weights using the EM Algorithm. This method has spread widely on the specialized literature, and albeit no strong theoretical optimality seems to exist, it enjoys adequate performance in Meteorological and Climatological applications.


Bayesian model averaging is used for combined estimation in $111$ articles (around $19\%$) throughout our revision. Albeit similar in purpose with combined prediction, we classified a work as a combined estimation article if its purpose was to estimate a common parameter to all models, but not a future observation. Combined estimators were employed to estimate a variety of quantities that might be appropriately modeled by plenty of models, such as population size \citep{King2001}, toxicity in clinical trials \citep{Yin2009}, breeding values in genetics \citep{Habier2010} and the probability of an economic recession \citep{Guarin2014}.

Conceptual discussions and methodological articles ammounted for $65$  data points (around $11\%$ of the dataset). This category presents a clear heterogeneity, as it comprises theoretical and conceptual advances in many directions. Articles in this category are, however, very similar with respect to its purpose to introduce an application of BMA to an existing problem or extend BMA to overcome limitations in some settings.

The former articles refer to seminal theoretical works like \cite{Raftery1996}, that introduced the BIC approximation to the Bayes factor and paved the way for many subsequent works. There were also pioneer works discussing the introduction of BMA to applied fields, like the methodology discussed in \cite{Fernandez2001} for variable selection in economical applications. Its use of Zellner's $g-$prior \citep{Zellner1986} for the regression coefficients (allowing for explicit model evidences) and MC3 composition for model space search were quickly adopted by many authors in very diverse economical studies. The introduction of BMA to the problem of selecting network structures in Bayesian Networks \citep{Friedman2003} also had a great impact in the Machine Learning literature, spawning a wealth of approximations for the ideal Bayesian averaged network. Finally, there were also novel developments in BMA that made it usable within a field. For instance, the variational Bayes approximation for the evidence proposed in \cite{Friston2007} sparked plenty of BMA applications to neurological datasets.  On the other hand, the latter articles deal with extensions of BMA to different Bayesian applications, like the BMA under sequential updating discussed in \cite{Raftery2010}.


The last category we employed to classify articles on usage pertains to review articles on BMA or the applications of model averaging to specific fields or models. There are $18$ such articles in our dataset, amounting to less than $3\%$ of the total. Some of the revision articles sparked the application of BMA in general like the seminal works by \cite{Hoeting1999} and \cite{Wasserman2000}, whereas more specific revisions exposed the methodology in other fields like Economics \citep{Geweke1999}, Genetics \citep{Fridley2009} and Physics \citep{Parkinson2013}. There are also revisions on model selection \citep{Kadane2004} and model uncertainty \citep{Clyde2004} in which BMA figures as a technique.

\subsubsection{Field of application}

As stated in Section \ref{SS:Field}, we classified the dataset into four main categories to give a broad idea of the application of BMA in  different fields. We divided our dataset into four categories regarding applications in the Biological and Life Sciences, Humanities and Economics, Physical Sciences and Engineering and Statistics and Machine Learning, respectively in order to infer if there was an increased penetration of BMA in either field.

\begin{figure}[hbtp]
\centering
\includegraphics[width=\linewidth]{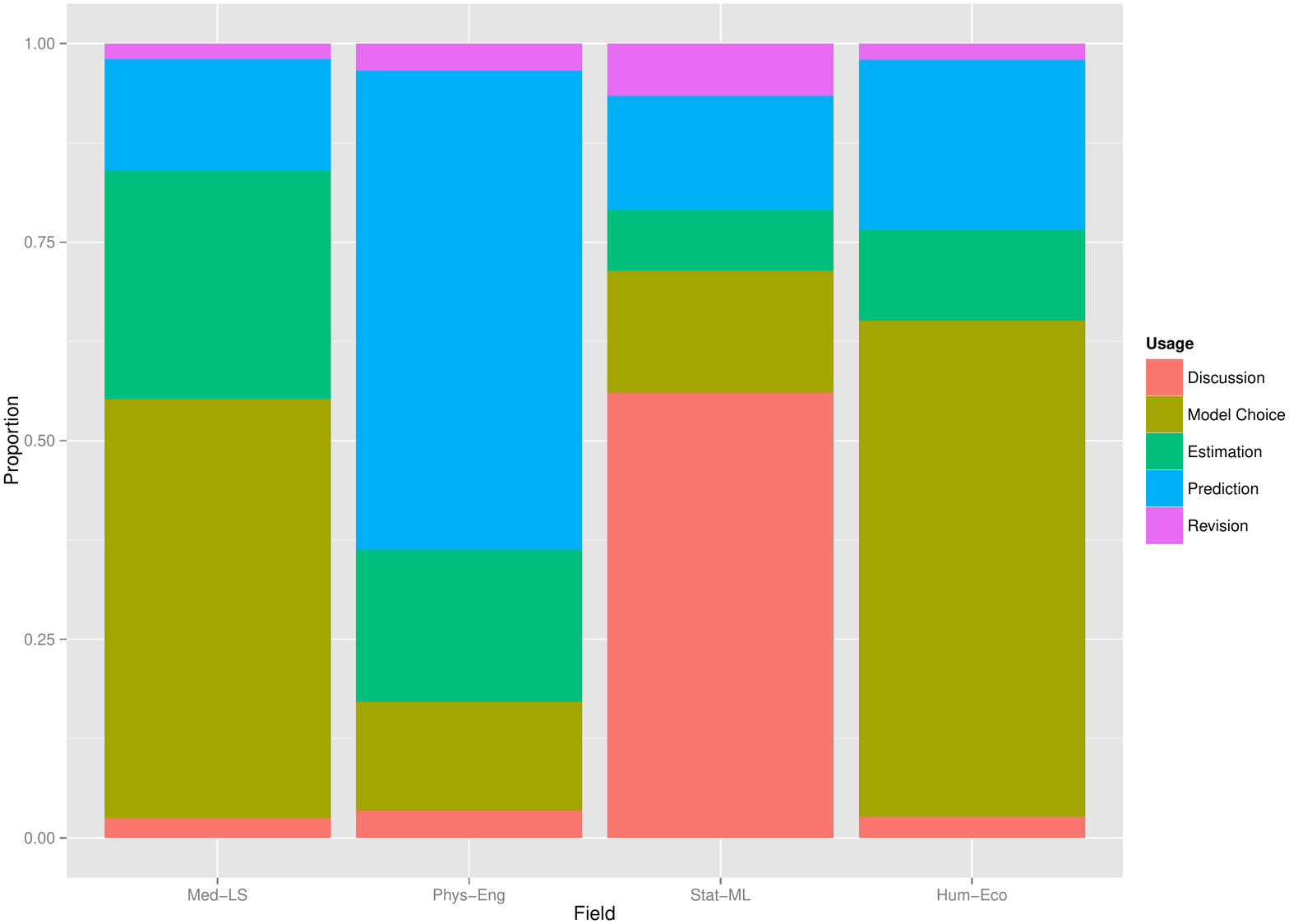}
\caption{Usage of BMA by field of application}
\label{F:UsoXArea}
\end{figure}


The field with most articles was Life Sciences and Medicine, with $201$ works corresponding to $34.24\%$ of all reviewed publications. Within these publications, there are some trends of note. In the medical sciences, BMA was used mostly for model choice purposes, as in variable selection of factors associated with false positives in diagnostic kits \citep{Ranyimbo2006}, weight loss \citep{Phelan2009}, leukemia progression \citep{Oehler2009}, structure in Bayesian networks for patient-specific models \citep{Visweswaran2010} and the combined estimation as in \cite{Yin2009}. BMA was also used extensively in Ecology, with applications to landscape ecology and geographical information systems \citep{Barber2006},  prediction of species distribution \citep{Thomsom2007}, capture-recapture models \citep{Arnold2010} and evolution \citep{Silvestro2014}.

The massive number of genetic variables obtained from biomarkers like microarrays and Single Nucleotide Polymorphisms (SNPs) that relate with observed characteristics through complex interactions fostered a rich Bayesian model selection literature in Genetics. BMA was employed to combine and select genetic information from metabolic pathways \citep{Conti2003}, quantitative trait loci \citep{Boone2008}, candidate SNPs for lung cancer risk \citep{Swartz2013} and a revision of Bayesian model selection models was performed in \cite{Fridley2009}. There is also the seminal work by \cite{Meuwissen2001} that introduced a class of SSVS-like estimation procedure in quantitative genetics that gained much traction in the Animal and Plant sciences literature, concerned with the prediction of genetic merit for selection \citep{Kizilkaya2010,Boddhireddy2014}. There also seems to be an increase of model averaging in neuroscience using the methods introduced by \cite{Penny2010}.


The Humanities and Economy field had $149$ articles ($25.38\%$ of the dataset), with the overwhelming majority being in economical applications. The earliest application to economics in our dataset is the revision of Bayesian modeling by \cite{Geweke1999}. The most adopted framework for model selection using BMA was introduced in \cite{Fernandez2001}, describing conjugate priors and a model search procedure that enjoys a broad application to many economical questions to this date, although were also articles applying the framework introduced by Adrian Raftery and collaborators through the R package this group developed like done in \cite{Goenner2010} and applications of SSVS \citep{Vrontos2008}. Forecasting using BMA in economical settings was discussed in \cite{Koop2004} and was adopted by many subsequent articles. There were also some occasional applications to  Political Science mostly on variable selection as in trade interdependence in conflict \citep{Goenner2004}  and some forecasting, like the study of the $2000$ american presidential election in \cite{Sidman2008}.

The application of BMA in the Physical Sciences and Engineering amounted for $146$ articles, about $24.82\%$ of the total. Most of this field was made of applications to forecasting problems in Meteorology and Climatology stemming from the seminal work of \cite{Raftery2005} in the field, using the ideas of BMA into an ensemble forecast with posterior model probabilities estimated using the EM algorithm. There were also some works in Physics, like the study of dark matter in \cite{Liddle2006} and the revision of BMA methods in astrophysics in \cite{Parkinson2013}.


Finally, there were $92$ articles in the Statistics and Machine Learning field. There are some methodological articles, like \cite{Raftery1996} on the BIC approximation and revisions on BMA like the highly influential \cite{Hoeting1999} and \cite{Wasserman2000} and model uncertainty \citep{Kadane2004,Clyde2004}. There were also some developments in model choice to more statistically sophisticated models like Generalized Linear Models (GLMs, \cite{Morales2006}), transformations \citep{Hoeting2002} and Generalized Auto-Regressive Conditional Heteroskedasticity (GARCH) models \citep{Chen2011}. In the Machine Learning literature, BMA was mostly used along with Bayesian Networks \citep{Friedman2003}, with theoretical developments allowing for the direct estimation of the averaged network without the necessity of the usual estimation procedure of estimating and combining all posteriors \citep{Dash2004,Cerquides2005}.

Since our revision spans through very diverse fields with very different questions, we investigated how these research questions were reflected in usage throughout the fields. A graphical summary can be observed in Figure \ref{F:UsoXArea}. Combined prediction is more common in the Physical Sciences field, which is mostly explained by the abundance of prediction articles using the ensemble method by \cite{Raftery2005}. Economics research questions seemed to focus more on the search for influent factors and determinants, leading to model and variable selection. Since economics dominated the Humanities and Economics field, one can observe a large number of model choice papers. Model selection was also very present in the Life Sciences and Medicine field, with more than half the revised articles. There is, however, an interest in combined estimation that is larger than other fields. Finally, most of the Statistics and Machine Learning literature revised concerns methodological advancements and discussions.

\subsubsection{Model priors}

Prior elicitation is a current and open research subject in Bayesian inference as a whole, with distinct currents advocating completely subjectivist researcher-driven prior distributions to completely agnostic and data-driven priors and many types of compromise in between. As discussed in Section \ref{S:Background}, this problem is compounded in BMA settings, as the space of the conceivable models is a more abstract parametric space, thus harder to measure in terms of a probability measure.

The most common answer to this uncertainty is to assume an uniform distribution over the model space through a vague prior. Such practice is, by far the most common and is adopted by more than $50\%$ of the revised articles, in $297$ publications. Given a finite number of $K$ models, the authors simply assume $\pi(M_l)=\frac{1}{K}$ for $l=1,\ldots,K$ or an equivalent formulation like assuming a Bernoulli prior with prior inclusions equal to $\frac{1}{2}$ in SSVS. In the case of an infinite number of models like in mixture problems, some authors adopt the alternative proposed in \cite{Richardson1997} of fixing a maximum number of classes and assuming an uniform distribution over the restricted model space.

Some authors circumvent the problem by adopting more traditional prior distributions verbatim from the literature. This practice was present in $98$ articles, around $16.6\%$ of all publications and there seems to be a few trends. Some of this adoption is caused by convenient conceptual frameworks - many authors in the economics literature simply adopted the prior, evidence estimation and model search proposed by \cite{Fernandez2001} verbatim. With the diffusion of BMA software first by personal request to Adrian Raftery's group with the S-Plus codes could be obtained by sending an email titled ``send BMA from S'' to the authors and then by the use of the \texttt{bma} R package, less attention was paid to the priors as many authors simply used whatever was default in these implementations.

We also observed $63$ articles that tackled the problem and derived model priors for posterior BMA. Some noteworthy examples in the literature include \cite{Medvedovic2002} that elicited model priors by the expected behavior of the cell cultures under study, the analytical considerations of the Bayesian networks of interest in \cite{Friedman2003}, the combination of multiple stakeholders in \cite{Mantyniemi2013} and the proposal of a cross-model correlations to elicit model priors in \cite{Garthwaite2010}.

The remainder $130$ articles amount to those that did not specify or apply any model priors. Albeit odd by Bayesian standards, these articles still apply BMA, but priors are not specified. The most common reason in our dataset is of the applications of the ensemble forecast method introduced in \cite{Raftery2005}, that just estimates a ``posterior'' weight using the EM algorithm with no mention of priors. There is also the adoption of default options in software packages for BMA for which we tracked down the default settings when possible, but that was not always possible.

\subsubsection{Evidence estimation}

The estimation of the marginal likelihood leading to the posterior model probability presents a considerable difficulty to be overcome in the applications of model averaging. As datasets get more complex, thus requiring more complex models, we aimed to investigate how the BMA literature deals with the problem of estimating the model evidence.

The integral in equation \eqref{E:evidence} is not available in closed form for most likelihoods besides in the (generalized) linear regression model with conjugate priors. These models are however, very popular and its widespread adoption explains most of the $232$ articles we classified as ``not applicable'' (NA) in our CCS. 

With respect to approximations, most revised articles approximated the evidence by analytical means, as done in $190$ articles. Although inferior to the Laplace approximation in theory \citep{Kass1995, Raftery1996} and simulations \citep{Boone2005}, most articles use the BIC approximation, as its value is given by most software available for parameter estimation in generalized linear models, making for a straightforward application of BMA. Its immediate availability from the maximum likelihood estimates, combined with some theory and available software for a wide class of GLMs justify its popularity. Its popularity also makes it the most widely misused approximation.  We encountered ``approximations''  based on other information criteria as a way to improve over the BIC, frequently using the Akaike Information Criteria (AIC), the Deviance Information Criteria (DIC) and other more \textit{ad hoc} information criteria. 

 Markov Chain Monte Carlo methods were used to estimate the evidence in $113$ articles.  The vast majority of there works use RJMCMC or SSVS methods, and as such, the posterior model probabilities are estimated by the sample average of an indicator variable for each model. There were, however, plenty of articles that employed the Importance Sampling approach by \cite{Gelfand1994} and the much criticized harmonic mean estimator \citep{Newton1994}, as they are simpler to implement in complex models than the trans-dimensional proposals and transformation required for RJMCMC. Still in spirit with MCMC but using a different technique, $17$ articles used the ratio of densities \citep{Chib1995} approach.

We also encountered $36$ articles in which the evidence was estimated through the use of Monte Carlo integration techniques (using independent samples, in constrast with the dependent Markov Chain samples used in MCMC), but there was no unifying trend over the practice. Most authors used importance methods, like Sampling-Importance-Resampling (SIR), in which a size $n$ sample is taken from the prior distributions and used to construct importance weights using the likelihood. Posterior samples are them obtained from resampling $m < n$ values with replacement using the constructed weights. With these samples, posterior inference and evidence estimation are direct even for complex models, like the stochastic differential equation model used in \cite{Bunnin2002}.

\subsubsection{Dimensionality}

As datasets get more massive and models more complex to aggregate different sources of information, it is common that large model spaces emerge. For instance, in the Genomics literature, the development of high throughput marker chips generated millions of Single Nucleotide Polymorphism (SNP) variables one desires to associate with a response variable inducing a dimensionality problem. As mentioned in Section \ref{S:Dim}, we investigated two approaches: previous filtering and stochastic search through Markov Chains.

Between these two approaches, stochastic search was the most prevalent one, with $144$ articles. Among these papers, there are three main trends. With the abundance of BMA in regression models, plenty of the literature was concerned with variable selection, which led to the widespread application of dedicated methods like SSVS that allow for a direct measure of association strength through the marginal posterior inclusion probabilities. A typical example of SSVS can be encountered in \cite{Blattenberger2012} applied to the search for risk factors related to the use of cell phones while driving. The Markov Chain Monte Carlo Model Composition (MC3, \cite{Madigan1995}) was also widely applied in variable selection settings, mainly in the economics literature following the seminal work by \cite{Fernandez2001b}. MC3 in turn is very similar in spirit with the Reversible Jump MCMC methods, which were mostly employed in our revision to choose between complex models instead of variable selection, like done in \cite{Wu2011}.

Some authors chose to perform a prior dimensionality reduction in the model space and then apply the model averaging, as done in $103$ articles in our dataset, most of them on variable selection for regression models. Among these prior reductions, the most popular was the Occam's Window criterion \citep{Madigan1994}, probably due to its implementation in the \texttt{bma} R package.  In some works, some  applied the Leaps and Bounds algorithm \citep{Furnival1974} prior to the Occam's Window to select highly likely variable subsets without the need of an exhaustive calculation.

However, there were cases where authors performed exhaustive calculations for all possible models and many cases where the number of models under consideration were small, leading to the $341$ articles marked as ``not applicable'' (NA) in our dataset. 

\subsubsection{Markov Chain Monte Carlo methods}

We investigated the application of MCMC methods throughout our dataset. For this particular classification, we excluded $24$ works that either presented purely revisions or conceptual discussions and did not apply any estimation methodology and articles in which BMA was applied but poorly described to the point we could not tell whether MCMC methods were used anywhere. The remainder $563$ articles were classified to their usage of MCMC or lack thereof. In our dataset, $261$ articled applied MCMC methods, whereas $302$ did not.

\begin{figure}[hbtp]
\centering
\includegraphics[width=\linewidth]{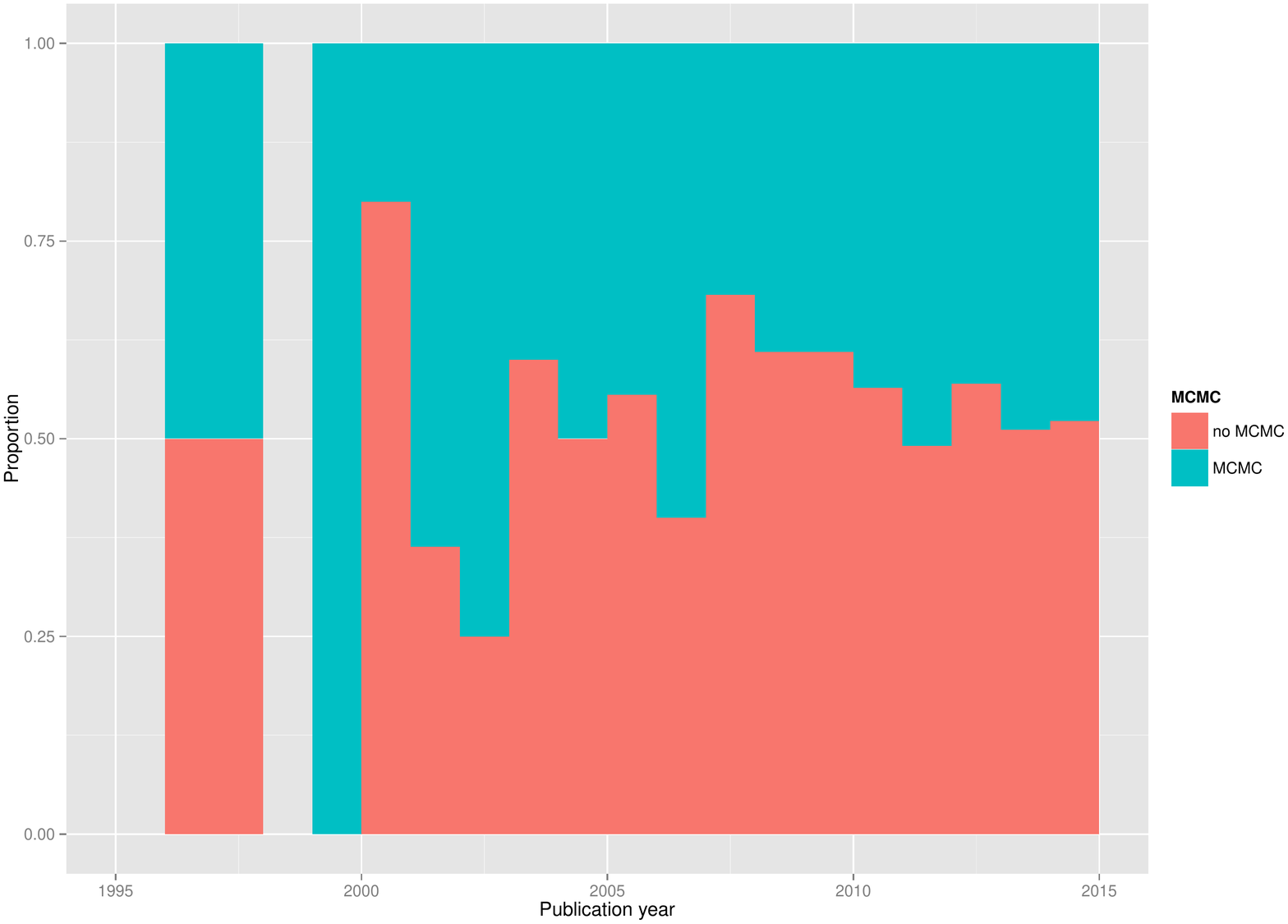}
\caption{Markov Chain Monte Carlo methods usage in the dataset}
\label{F:MCMC}
\end{figure}

The temporal pattern of MCMC usage, separated by publication year can be found in Figure \ref{F:MCMC}. Although we suspected that MCMC would become more popular after the popularization of freely available software, it seems like MCMC methods were not as prevalent in the literature as expected and its usage did not increase much even with more available software and cheaper computational power. MCMC was applied in approximately half the articles published in any given year with the exception of the early years of the $2000$s.

\subsubsection{Simulation studies}

Our dataset was also classified with respect to the presence of simulation studies. Overall, $19$ articles were not clear on their usage of simulations, and were therefore excluded from this classification. Of the remaining articles, the majority composed of $375$ articles did not perform any form of simulations, whereas $193$ did some sort of simulation study.

\begin{figure}[hbtp]
\centering
\includegraphics[width=\linewidth]{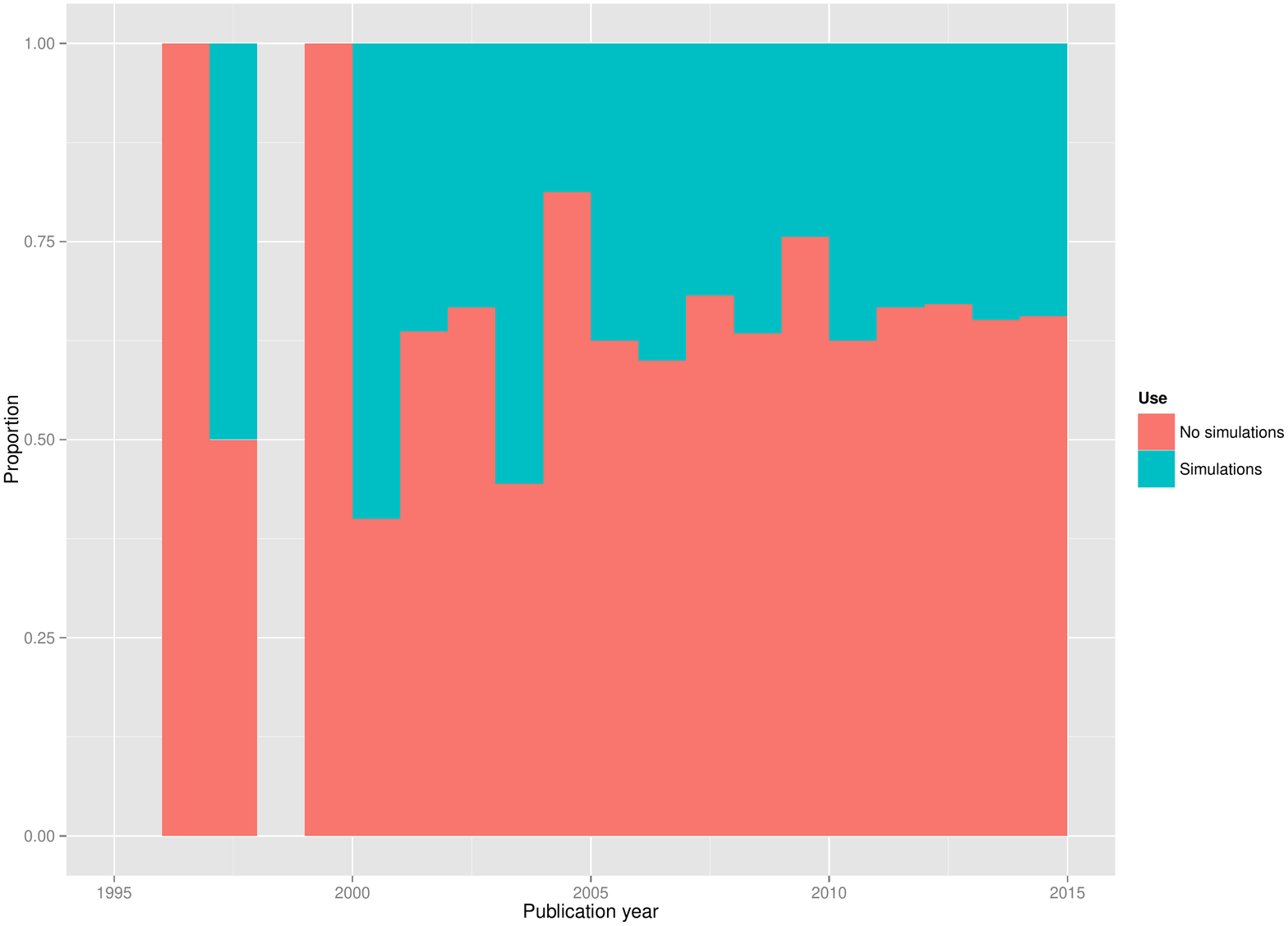}
\caption{Proportion of simulation studies}
\label{F:Sims}
\end{figure}

A more detailed classification by year of publication can be encountered in Figure \ref{F:Sims}. As in the MCMC classification, the increased computational power did not seem to have an effect on the realization of simulation studies. One can also observe that simulations seem to be unpopular overall, never accounting for more than half the published papers in any given year.

\subsubsection{Data-driven validation} 

We classified the literature on the use of data-driven validation of the results obtained by model averaging. The majority of articles ($358$ papers) did not perform any validation whatsoever.  Among the articles that performed some kind of data-based validation, the overwhelming majority of $179$ articles used cross-validation by splitting the dataset into two disjoint subsets, fitting the desired models to one of the subsets and validating their performance on the other. The slightly more sophisticated K-fold cross-validation was only applied in $17$ articles.

Curiously, even though BMA is a Bayesian technique, there was only a single application of posterior predictive checks \citep{Barber2006}, a data-driven Bayesian goodness-of-fit criteria based on the posterior distribution deriving conclusions rooted in the Bayesian paradigm.

\section{Concluding remarks}\label{S:Conc}

This work performed a methodical literature review of Bayesian Model Averaging studies in the $1996-2014$ time period. After a thorough search, we employed a conventional content analysis, and proposed a novel conceptual classification scheme that we  applied to the literature, revising and classifying $587$ articles from a large variety of publications spanning a wide range of applications.

Although much was discussed in the formulation of the proposed classification scheme and its results when applied to the selected literature, some limitations to the present work still exist. First, our search was limited to peer-reviewed articles published in digitally available periodicals. As such, many developments in BMA possibly published in conference proceedings, dissertations or theses and books might have been overlooked. Secondly, said periodicals were restricted to the titles listed on the four databases mentioned in Section \ref{S:LitSearch} and, albeit an effort was made to be as inclusive with the literature as possible, non-listed titles were not included except when cited in selected articles and passed the exclusion criteria. Finally, the search was initiated through specific queries on these databases, and as such, relevant articles might have been overlooked when we restricted terms.

Limitations aside, this revision can provide relevant insights into the current BMA literature and some indications of future developments. With no aspiration to be exhaustive, we point out a few. Namely, we have five observations:

\begin{enumerate}
\item The methodology provides for a very flexible account of model uncertainty which can in principle be applied to any problem, but not much was done in model choice using BMA aside from variable selection in regression models, with the exception of BMA in Bayesian Networks present in the Machine Learning literature of the early $2000$s.
\item Not much was done in methodological developments in the statistical literature aside from the seminal works in the late $1990$s, limiting its application mostly to (generalized) linear regression models in spite of more complex models being proposed in the same time period. For instance, the problem of evidence estimation is still either circumvented by using convenient likelihoods and conjugate priors or solved through the BIC approximation, which, being reliant on plenty of regularity conditions, might not be adequate for more complex models. Much of the developments in MCMC after the textbook algorithms (i.e. straightforward applications of the Gibbs's Sampler and Metropolis-Hastings algorithms) were also mostly absent from our dataset. 
\item There was no significant discussion on computational costs of BMA. There are distinct computational problems derived from dealing with very large datasets, very large and complex model spaces and in the fit of complex models that were not approached in the revised literature.
\item The vague prior is still the most used model prior and, albeit convenient it might not be the best choice for all problems. Not much was done in prior elicitation or the reference priors for BMA. 
\item Neither simulation studies or validation methods were popular in the literature. This presents a very serious issue, as BMA usually deals with many models drawing from sometimes very distinct assumptions and therefore reaching distinct conclusions. Using BMA without any validation might lead to an overconfidence in the conclusions, the very problem model averaging is proposed to mitigate.
\end{enumerate}

\bibliography{BMA}
\bibliographystyle{dcu}

\end{document}